\documentclass[pre,twocolumn,usletter,showpacs,superscriptaddress]{revtex4}

\usepackage{amsmath}
\usepackage{amsfonts}
\usepackage{amssymb}

\usepackage [ansinew] {inputenc}
\usepackage[dvips]{graphicx}
\usepackage{natbib}

\begin{document}
\author{Vladimir Garc\'{\i}a-Morales}
\email{vmorales@ph.tum.de}
\affiliation{Physik-Department E19,
Technische Universit\"{a}t M\"{u}nchen, James-Franck-Str. 1,
D-85748 Garching, Germany}
\author{Julio Pellicer}
\author{Jos\'{e} A. Manzanares}
\address{Departament de Termodin\`{a}mica,
Universitat de Val\`{e}ncia, C/Dr. Moliner 50, E-46100 Burjassot,
Spain}

\title{Thermodynamics based on the principle of least abbreviated action:
entropy production in a network of coupled oscillators}

\begin{abstract}
\noindent We present some novel thermodynamic ideas based on the
Maupertuis principle. By considering Hamiltonians written in terms
of appropriate action-angle variables we show that thermal states
can be characterized by the action variables and by their
evolution in time when the system is nonintegrable. We propose
dynamical definitions for the equilibrium temperature and entropy
as well as an expression for the nonequilibrium entropy valid for
isolated systems with many degrees of freedom. This entropy is
shown to increase in the relaxation to equilibrium of macroscopic
systems with short-range interactions, which constitutes a
dynamical justification of the Second Law of Thermodynamics.
Several examples are worked out to show that this formalism yields
the right microcanonical (equilibrium) quantities. The relevance
of this approach to nonequilibrium situations is illustrated with
an application to a network of coupled oscillators (Kuramoto
model). We provide an expression for the entropy production in
this system finding that its positive value is directly related to
dissipation at the steady state in attaining order through
synchronization.
\end{abstract}
\pacs{45.50.Jf, 05.70.Ln, 05.45.Xt} \maketitle \pagebreak

\section{Introduction}

Variational principles occupy a privileged place in the history of
physics. They provide approximate results to formidable problems
where it is difficult to find analytical solutions from other
methods \cite{Gray04}. The first variational principle was
established in mechanics by Maupertuis (1744). Its interest and
importance was later realized by Lagrange, who referred to it as
\emph{the most beautiful and important discovery of Mechanics}
\cite{Gray04}. This principle led to an elegant geometrization of
mechanics \cite{Landau82, Casetti00} for conservative systems, and
had enormous consequences in the subsequent development of
physics.

The variational principles of mechanics constituted the key idea
of the former mechanical approaches to thermodynamics. These were
first attempted independently by Boltzmann and Clausius and later
continued by Helmholtz \cite{Bailyn94,Gallavotti99}. These
pioneering works allowed for a certain understanding of
thermodynamics from mechanics \cite{Gallavotti99, Campisi05} but
their importance was later underestimated in favour of a
statistical interpretation of thermodynamics. The history of these
developments and the confusion surrounding Boltzmann ideas have
been recently illuminated by some authors \cite{Gallavotti99,
Lebowitz, Lebowitzb, Cohen05, Gross, Gallavotti95}.

Despite one century of successful results coming from the
statistical approach, the bridge between microscopic dynamics and
macroscopic behaviour is still an open issue. Probabilistic
concepts are not only extremely useful in practical situations but
play also a central role in laying the microscopic foundations of
Thermodynamics. The following criticism by Einstein to Boltzmann
(who realized the importance of the statistical approach in later
works) \cite{EGDCohen} still holds: \emph{the systems considered
in statistical mechanics are dynamical systems, consisting of
moving particles [and therefore] the statistical results should
all be derivable from the dynamics}.

In this article, we present some novel thermodynamic ideas which
are grounded in the Maupertuis principle as well as in the modern
general theory of dynamical systems. Ergodicity and probability
assumptions (phase space averaging) are not invoked, and only time
averages coming from the microscopic mechanics are considered. Our
approach makes use of Hamiltonians given in terms of action-angle
variables
\begin{equation}
H=H_{0}(\mathbf{J})+H_{1}(\mathbf{J},\mathbf{\theta})
\label{Hamiltonian}
\end{equation}
where $(\mathbf{J},\mathbf{\theta})\equiv(\{J_{i}\},
\{\theta_{i}\})$ are vectors containing suitable action and angle
variables respectively, and $H_{0}$ and $H_{1}$ are, respectively,
the ''integrable'' and ''nonintegrable'' parts of the Hamiltonian.
It is generally possible to construct such Hamiltonians from, for
example, mechanical variational principles \cite{Gray04,
Percival74, Kaasalainen}. Then, we show that thermal states can be
fully characterized by the action variables. When the system is
nonintegrable, these variables do not need to be adiabatic
invariants but the entropy, defined as a function of them, must
not change significantly within a period of the oscillation of any
degree of freedom (DOF).

The outline of the paper is as follows. In Section II we introduce
the dynamical concepts of equilibrium temperature, entropy and
heat differential. The central concept in our approach is the
nonequilibrium entropy for a system of many DOFs, which is
introduced in Section III. We then link this result to the ones in
Section II and a natural, dynamical definition of probability
arises, yielding an Einstein relationship at the thermodynamic
limit. In Section IV we show that entropy attains its maximum at
equilibrium for macroscopic systems with short-range interactions.
This connects results in Sections II and III. In Section V we
apply this dynamically based thermodynamic formalism to several
systems of different nature: ideal systems (noninteracting
oscillators, ideal gas) and systems with short and long-range
interactions. We show how our dynamical definitions correctly
reproduce previous equilibrium results obtained from the
microcanonical ensemble. Finally, in Section VI, we provide an
application to nonequilibrium stationary states in a network of
coupled oscillators (Kuramoto model). We study the entropy
production in this system by means of the nonequilibrium theory
developed in Sections III and IV and illustrate its physical
significance. We find that the entropy production at the
stationary state is directly related to dissipation in attaining
order far from equilibrium through synchronization. Numerical
calculations are performed to clarify this result.

\section{Equilibrium dynamical temperature and entropy}
In the prehistory of Statistical Mechanics (i.e. before the
introduction of any probabilistic concept) Boltzmann provided a
mechanical foundation of thermodynamics for monocyclic systems
with only one DOF \cite{Boltzmann,Bailyn94}. The constant energy
trajectories in phase space of these systems (e.g. a single
harmonic oscillator) were bounded, periodic orbits, and Boltzmann
defined heat as the energy difference between two orbits. Then he
introduced a dynamical temperature and proved that its reciprocal
is an integrating factor for the heat differential, which allowed
him to identify the entropy of the system. In this article we
follow a different path inspired in the Maupertuis principle. Our
approach is valid for $f$ DOFs and reduces to Boltzmann's theory
for $f=1$.

Consider a system whose Hamiltonian $H(\textbf{q},\textbf{p},t)$
depends on time $t$ and generalized position coordinates
$\textbf{q}(t)=(q_{1}(t),...,q_{f}(t))$ and momenta
$\textbf{p}(t)=(p_{1}(t),...,p_{f}(t))=\triangledown_{q}
\mathcal{A}$, where $f$ is the number of DOFs. The Hamilton-Jacobi
equation reads then  \cite{Landau82}
\begin{equation}
\frac{\partial \mathcal{A}(\textbf{q},t)}{\partial
t}+H(\textbf{q},\textbf{p},t)=0 \label{HJ}
\end{equation}
where $\mathcal{A}=\int Ldt$ is the Lagrangian action and $L$ is
the Lagrangian of the system. For conservative systems, $H=E=const.$, action can be varied with
respect to the time $\tau$ of the trajectory and the
following equation is satisfied \cite{Landau82}
\begin{equation}
\delta \mathcal{A}+E\delta \tau=0 \label{variationTime}
\end{equation}
Since time can be separated in the Hamilton-Jacobi equation, after
integration we obtain $\mathcal{A}=J_{E}- E\tau$ where
$J_{E}=\int_{\textbf{q}_{0}}^{\textbf{q}_{1}}
\textbf{p}d\textbf{q}$ is the abbreviated action. The Maupertuis
principle of classical mechanics states that \emph{the abbreviated
action $J_{E}$ of the trajectory of the motion is an extremum over
all possible energy-conserving trajectories that pass through
ending points $\textbf{q}_{0}$ and $\textbf{q}_{1}$ in an
arbitrary time $\tau$} \cite{Landau82}.

From the total variation of the action $\delta
\mathcal{A}=\frac{\partial J_{E}}{\partial E}\delta E-\tau\delta
E-E \delta \tau$, and Eq. (\ref{variationTime}), we have
\begin{equation}
\frac{\partial J_{E}}{\partial E}=\tau \label{time}
\end{equation}
In the case of closed orbits this equation is written as
\begin{equation}
\frac{\partial J_{c}}{\partial E}=\tau_{c} \label{time closed orbit}
\end{equation}
where $J_{c}=\oint\textbf{p}d\textbf{q}$ is the abbreviated action
of a closed orbit and $\tau_{c}$ is the (Poincar\'e) time needed
for its completion. It is then clear that the product of $J_{c}$
and the recurrence  frequency $\omega_{c}\equiv 1/\tau_{c}$ has
dimensions of energy. We define the \emph{dynamical} equilibrium
temperature as
\begin{equation}
T^{(eq)} \equiv \frac{\omega_{c}J_{c}}{fk} \label{temperature}
\end{equation}
where $k$ is the Boltzmann constant. A similar definition was
introduced by Boltzmann for the case $f$ = 1 (see \cite{Bailyn94},
p. 419). Since $J_{c}$ is an invariant of the motion and does not
depend on the choice of coordinates \cite{Landau82}, the
equilibrium temperature does not depend on this choice either.

The dynamical definition of the equilibrium entropy $S^{(eq)}$ is
introduced so that the thermodynamic equation $\partial S^{(eq)}
/\partial E=1/T^{(eq)}$ is satisfied.  Since Eqs.(\ref{time closed
orbit}) and Eq. (\ref{temperature}) lead to
\begin{equation}
fk\frac{\partial \ln J_{c}}{\partial E}=\frac{1}{T^{(eq)}}
\label{entropydif}
\end{equation}
the equilibrium entropy $S^{(eq)}$ is defined as
\begin{equation}
S^{(eq)} \equiv fk \ln \left(a J_{c}\right) \label{entropy}
\end{equation}
where $a$ is an integration constant with dimensions of inverse
action (in the following we take $a$ as unity). Consistently, the
heat differential is defined as
\begin{equation}
dQ \equiv \omega_{c} dJ_{c}
\end{equation}
which represents \emph{the variation of energy involved in
changing the action on the closed trajectory}. Thus, it is
satisfied that $1/T^{(eq)}$ is an integrating factor for the heat
differential, $dQ =T^{(eq)}dS^{(eq)}$.

These thermodynamic expressions are useful if we are able to calculate
 $J_{c}$ from the microscopic dynamics. For systems with $f=1$ and bounded
orbits this is always possible since the Hamiltonian is
integrable. The constant energy $E$ can then be directly related
to $J_{c}$. For systems with many DOFs, ergodic theory has
provided the means for calculating $J_{c}$. This theory
establishes the equality between time averages and phase space
averages under certain conditions and the Birkhoff theorem. Our
approach does not require ergodicity but leads to the same results
if Birkhoff theorem holds, as shown in Appendix A.

\section{Nonequilibrium entropy}
\label{NoneqSection}

To extend the formalism to nonequilibrium systems with many DOFs,
we write the Hamiltonian as in Eq. (\ref{Hamiltonian}) by means of
an appropriate canonical transformation. This yields a set of
action-angle variables $(\mathbf{J},\mathbf{\theta})$ whose
evolution is then considered. The term $H_{1}$ in Eq.
(\ref{Hamiltonian}) does not come from a perturbation expansion
and it is not necessarily smaller than $H_{0}$. $H_{1}$ is
associated to energy exchange between DOFs in the form of heat. In
that case the $J_{i}$'s are neither adiabatic invariants nor
constants of the motion but suitable generalized coordinates with
dimensions of action. These action-angle variables are to be
understood as effective ones, coming from approximations
consistent with the dynamics of the Hamiltonian under
consideration (a criterion to be satisfied for appropriate action
variables is given below). Since each DOF oscillates periodically
with its effective frequency, it is physically meaningful to
consider the action $J_{i}$ of the DOF $i$ on its closed cycle
with the understanding that this cycle can be completed by the DOF
going back and forth following the Hill's region (i.e. the region
in configuration space to which the DOF is confined) until it
reaches its initial angle. Only in the limit of vanishing $H_{1}$
are the action variables adiabatic invariants also: they
correspond to the integrable part of the Hamiltonian and solve
\emph{exactly} the integrable dynamics. This kind of
representations is used in plasma physics and can be constructed
for general Hamiltonians \cite{Morrison}.

We define the "temperature" and "entropy" of the DOF $i$ as
\begin{equation}
T_{i} \equiv \frac{\omega_{i}J_{i}}{k} \label{temperature DOF i}
\end{equation}
\begin{equation}
S_{i}\equiv k\ln J_{i} \label{entropy DOF i}
\end{equation}
so that the thermodynamic formalism for one DOF resembles that for
the whole system. The energy of each DOF $\varepsilon_{i}$ ($i$ =
1, ... ,$f$) can vary subject to the constraint that the total
energy $E=\sum_{i=1}^{f}\varepsilon_{i}$ is constant, and it is
satisfied that $\partial S_{i} /\partial \varepsilon_{i}=1/T_{i}$.
These definitions are reasonable because Eqs. (\ref{temperature})
and (\ref{entropy}) apply to \emph{any} mechanical system,
regardless of its size, and each DOF in the Hamiltonian behaves as
a ''system'' with energy $\varepsilon_{i}$. However, contrarily to
the many-DOF system for which the total energy $E$ is a constant,
each DOF usually exchanges energy with the others (provided that
$H_{1}$ in Eq.(\ref{Hamiltonian}) is nonvanishing and the DOFs are
separated in $H_{0}$). The energy $\varepsilon_{i}$ is to be
understood as a \emph{time average} on the Hill's region in which
the DOF moves. This is clear from the Hamilton-Jacobi equation,
which can be applied both to the composite system and to each DOF
separately with the understanding that, in the latter case, each
DOF behaves as a system exchanging energy with the others and its
energy is, therefore, a time average.

The entropy of one DOF, Eq. (\ref{entropy DOF i}) can of course
change with time but to be physically meaningful it must not
change significantly during one period $\tau_{i}=1/\omega_{i}$ of
the oscilation of the DOF i. The DOF $i$ has explored its whole
available Hill region only after completing a period. The entropy
of one DOF is defined in terms of the \emph{whole} available Hill
region at a given time $t$ and it would be inconsistent not to
have an almost constant value of this quantity within one period
of the oscillation. This leads to stablish a criterion on the
suitability of the action variables which, as pointed out above,
need not to be adiabatic invariants. The condition to be satisfied
comes from expanding the entropy of a DOF $S_{i}$ in a time
$t+\tau_{i}$ respect to its value at time $t$. Keeping only linear
terms in $\tau_{i}$ we have
\begin{equation}
\frac{S_{i}(t+\tau_{i})}{S_{i}(t)} \approx
1+\frac{\tau_{i}}{S_{i}(t)}\frac{dS_{i}(t)}{dt} \label{crit1}
\end{equation}
The action variables are appropriate when the absolute value of
the second term in the r.h.s of Eq. (\ref{crit1}) is much lower
than unity. This implies
\begin{equation}
\left|\frac{1}{\omega_{i}}\frac{d}{dt}\ln \ln J_{i}\right| << 1
\label{crit2}
\end{equation}
If the action variables $J_{i}$ are adiabatic invariants this
criterion is automatically (and in the best way) satisfied. But
this is only a sufficient condition and not a necessary one. Given
any Hamiltonian in action-angle variables, the criterion in Eq.
(\ref{crit2}) can be applied to decide if the action variables are
appropriate or if a canonical transformation to find a better set
of variables is needed. If this condition is met, the temperature
of one DOF is also well defined from Eq. (\ref{temperature DOF i})
\begin{equation}
\frac{1}{kT_{i}} = \frac{\partial \ln J_{i}}{\partial H}
\end{equation}

It is to be noted that action variables bear geometric properties
of the phase space in the large \cite{Arnold78}. This can be seen
in the fact that in obtaining the action variable $J_{i} \equiv
\oint p_{i}dq_{i}$ the momentum of the $i$-th DOF $p_{i}$ is
integrated on a closed domain in configuration space involving the
coordinate position $q_{i}$. This closed domain (a 'monocycle') in
$q_{i}$ is the definition of Hill's region of DOF $i$. The Hill's
region can be a macroscopic object and, therefore, action
variables connect in this way the micro- and macroscales: The
former one is suggested by the fact that we address each DOF
separately; the latter comes in the distance over which the DOF
moves, an information contained in its action variable. In a
complicated problem, any assumption made in approximating action
variables is connected to an assumption on how the trajectory
fills the available phase space. This is of special relevance for
the Second Law (discussed in the next section) to hold since a
nonvanishing $H_{1}$ in Eq. (\ref{Hamiltonian}) is then required.
This in turn implies that the action-angle variables considered
are already 'effective' i.e. 'approximate' ones (the exact
action-angle variables would make the Hamiltonian independent of
angle variables, the system being integrable). A nonvanishing
$H_{1}$ in the action-angle representation is necessary to get
relaxation to equilibrium coming from the thermalization of the
DOFs. This term exist because of, for example, having interactions
between many DOFs that make the system nonintegrable. The
coarse-graining procedure as employed in the foundations of
standard Statistical Mechanics can be also thought as contained in
a particular assumption made on the action variables: the one
sketched in Appendix A in which every possible point in phase
space is equally visited and distributed over the Hill's regions
of all DOFs. Under such an approximation the dynamical trajectory
is \emph{a one-cycle permutation of all attainable points (or
'microstates') in phase space}. This allows to speak about a phase
space volume instead of a dynamical trajectory and to introduce
well-behaved probability distributions related to 'densities' of
occupancy of the phase space. In \cite{GarciaMorales06} this and
other assumptions leading to different Statistical Mechanics
formalisms are discussed from a different (but closely related)
viewpoint. Common to both lines of reasoning in
\cite{GarciaMorales06} and here is that for the possibility of
providing sound equilibrium statistical arguments motivated by the
dynamics, the system trajectory should be long enough in order to
approximate the volume of phase space in which it is contained.

A situation of upmost interest occurs when the total energy is
equally shared by all DOFs of the system, so that $\varepsilon_{i}
= \varepsilon^{(eq)} \equiv E/f$ for every DOF. In this case, the
temperature of every DOF is equal to the equilibrium temperature
of the system defined by Eq.(\ref{temperature}), $T_{i}=T^{(eq)}$.
This equilibrium state of the system is characterized by the
conditions $T_{i}=T^{(eq)}$ and $J_{c}\omega_{c}$ finite. In
relation to the latter condition, it must be noticed that
$\omega_{c}$ can be vanishingly small for macroscopic systems but
then $J_{c}$ is large. Thus, in an equilibrium state it is
satisfied that
$J_{c}\omega_{c}=fkT^{(eq)}=\sum_{i=1}^{f}J_{i}\omega_{i}$ and
therefore
\begin{equation}
fkT^{(eq)}=\sum_{i=1}^{f}J_{i}\frac{\partial{E}}{\partial{J_{i}}}=k\sum_{i=1}^{f}\frac{\partial{E}}{\partial{S_{i}}}
\label{energyeqi}
\end{equation}
This provides the energetic equation of the system.

In a nonequilibrium state, however, the temperatures and energies
of each DOF are different. The nonequilibrium entropy of the
system in such cases is
\begin{equation}
S \equiv \sum_{i=1}^{f}S_{i}=k\ln\prod_{i=1}^{f}J_{i} = fk\ln
J_{E}^{*} \label{superentropy}
\end{equation}
It depends on each $\varepsilon_{i}$ through the action variables.
This entropy corresponds to an open segment on the closed
trajectory with abbreviated action
$J_{E}^{*}\equiv\left(\prod_{i=1}^{f}J_{i}\right)^{1/f}$, and the
last equality in Eq. (\ref{superentropy}) makes this explicit. It
is worth noting that Eqs.(\ref{entropy DOF i}) and
(\ref{superentropy}) coincide with Eq. (\ref{entropy}) when the
system has only one DOF. The same applies to temperatures defined
by Eqs. (\ref{temperature}) and (\ref{temperature DOF i}).

For a closed trajectory it is satisfied that
\begin{equation}
\omega_{c}J_{c}=
\omega_{c}\int_{0}^{\tau_{c}}\mathbf{p\dot{q}}dt=<2K>_{\tau_{c}}
\label{kinetic closed}
\end{equation}
where $K$ is the kinetic energy. (It may be remembered that the
condition $\delta J_{E}=0$ leads to the equation of the geodesics
of the Jacobi metric $g_{ij}=2K\delta_{ij}$ \cite{Casetti00}.) A
segment of the closed trajectory is said to be typical if
\begin{equation}
<2K>_{\tau}=\frac{1}{\tau}\int_{t_{1}}^{t_{1}+\tau}\mathbf{p\dot{q}}dt
=<2K>_{\tau_{c}} = fkT^{(eq)} \label{tempopen}
\end{equation}
where $\tau$ is the duration of the segment (in all what follows
$\tau$ must be understood as a much longer time than the period
$\tau_{i}$ of the slowest DOF $i$). Therefore, the nonequilibrium
entropy for a typical segment is equal to the equilibrium entropy,
and this opens the possibility of defining the equilibrium
thermodynamic variables in terms of the action over a typical
segment rather than in terms of the action over the whole closed
(macro)trajectory, as done in the previous section. That is, we
can consider that at equilibrium the system moves along typical
segments. These considerations are further explored in Appendix B.
In relation to this, it is remarkable that the existence of an
equilibrium state implies, from Eqs. (\ref{entropy}) and
(\ref{superentropy})
\begin{equation}
\lim_{t\to
\tau_{r}<\tau_{c}}\left[\prod_{i=1}^{f}J_{i}(\varepsilon_{i})\right]=[J_{c}(E)]^{f}
\label{limiting}
\end{equation}
where $\tau_{r}$ is the relaxation time. That is, if the system
has a state of thermodynamic equilibrium \emph{it attains this
state after a time $\tau_{r}$ lower than $\tau_{c}$}. This
suggests that \emph{the lower is $\tau_{r}$ (compared to
$\tau_{c}$) the more accurate is the description of the
equilibrium state in terms of typical segments}. This is because
departures from equilibrium are then short and, therefore, the
closed trajectory contains a higher number of typical segments.

Since we can use Eq.(\ref{superentropy}) also to deal with
equilibrium situations provided we address typical segments,
action variables illuminate first-order equilibrium phase
transitions involving a discontinuous change in the entropy. If we
think on a Lennard-Jones fluid, for example, the Hill's region of
a DOF of a bound particle changes discontinuously when changing
(continuously) the total energy of the particle from $E<0$ to
$E>0$. In the former situation, the Hill's region is bounded by
the attractive region of the potential. In the latter one, the
Hill's region is bounded by the macroscopic volume where the fluid
is contained. This signals the transition from a liquid to a gas
state, involving a discontinuity on the value of the action
variables of many DOFs (since their energy is varied continuosly
but their Hill's regions change abruptly) and, therefore, a
discontinuity on the entropy, Eq. (\ref{superentropy}).

The probability $P(\gamma)$ that the system can be found in a
given segment $\gamma$ [i.e. a piece of micro-trajectory on the
closed (macro)trajectory] is given by the ratio between its
duration $\tau$ and the total duration of the closed trajectory
$\tau_{c}$
\begin{equation}
P(\gamma) = \frac{\tau}{\tau_{c}}=\frac{\partial
J_{E}^{*}/\partial E}{\partial J_{c}/\partial
E}=\frac{1}{fk}\frac{\partial S(\gamma)/\partial E}{\partial
J_{c}/\partial E}e^{S(\gamma)/fk} \label{probability}
\end{equation}
where $S$ is the nonequilibrium entropy in the segment. If we ask
for the probability of measuring certain macrovariables (as
$T^{eq}$) during a finite measurement time, it is clear from the
above that this is given by $P(\gamma_{1} \bigcup \gamma_{2}...
\bigcup \gamma_{n})=P(\gamma_{1})+P(\gamma_{2})+...P(\gamma_{n})$
where the $\gamma_{i}$'s are all segments on the closed trajectory
consistent with the macrostate characterized by the macrovariables
whose probability we ask. Two consecutive typical segments of
durations $\tau_{1}$ and $\tau_{2}$ produce a typical segment of
duration $\tau_{1}+\tau_{2}$. However, the contrary is not
necessarily true: not all pieces on an arbitrary decomposition of
a typical segment are also, in general, typical ones.

Eq.(\ref{superentropy}) suggests a roughly linear dependence of
$S$ on $f$ and, in the thermodynamic limit, Eq.
(\ref{probability}) can be simplified to
\begin{equation}
P(\gamma) \approx c e^{S(\gamma)/fk} \label{Einstein}
\end{equation}
where $c$ is a constant. This equation states that segments with
the highest entropy are also the most probable ones. These
segments are also the typical ones since, on them, the average in
Eq. (\ref{tempopen}) coincide with the average over the whole
trajectory. This is a result of the mean value theorem for
integrals. Note, finally, the similarity of Eq. (\ref{Einstein})
and Einstein's inversion formula \cite{Landsberg}. The latter
equation is the basis for Gibbs' variational principle (maximum
entropy approach) that allows to develop equilibrium statistical
thermodynamics.

\section{The Second Law in macroscopic systems with short-range
forces} \label{SecondLaw}

In most macroscopic systems with short-range interactions the
relation between the energy and the action for every DOF is
$J_{i}=c_{i}\varepsilon_{i}^{p}$ where $p$ is a positive exponent
(and $c_{i}$ does not depend on energy) and the entropy is then
\begin{equation}
S=k\ln \prod_{i=1}^{f}c_{i}+kp\ln \prod_{i=1}^{f}\varepsilon_{i}
\label{superentropy2}
\end{equation}
At equilibrium, $\varepsilon_{i}=\varepsilon^{(eq)} \equiv E/f$, $E=pfkT^{(eq)}$ from Eq. (\ref{energyeqi}), and
\begin{equation}
S^{(eq)}= k \ln \prod_{i=1}^{f}c_{i}+ fkp \ln \varepsilon^{(eq)}
\label{superentropyeq}
\end{equation}
We can use the following mathematical inequality \cite{Abramowitz}
(the geometric mean of a series of values $\varepsilon_{i}$ is
lower than its arithmetic mean)
\begin{equation}
\left(\prod_{i=1}^{f}\varepsilon_{i}\right)^{1/f} \le
\frac{1}{f}\sum_{i=1}^{f}\varepsilon_{i}=\varepsilon^{(eq)}
\label{unequal}
\end{equation}
then, since $p$ is positive, by comparing
Eqs.(\ref{superentropy2}) and (\ref{superentropyeq}) we obtain
that
\begin{equation}
S^{(eq)} \ge S \label{SecondLaw}
\end{equation}
where the equality only holds at equilibrium. Therefore, in the
evolution to equilibrium $\Delta S \ge 0$. This constitutes the
mathematical statement of the second law for conservative
macroscopic systems with short-range interactions.

Thermalization results from the coupling between DOFs and leads to
the homogeneity required for all thermal properties at
equilibrium. Obviously, no such coupling is possible in a
conservative system with $f=1$, and this is always ``at
equilibrium". During the relaxation to equilibrium, each $J_{i}$,
$\varepsilon_{i}$, $S_{i}$ and $S$ can change with time. Whether a
system has a state of thermodynamic equilibrium or not is
determined by the form of the nonintegrable part
$H_{1}(\mathbf{J},\mathbf{\theta})$ of the Hamiltonian in Eq.
(\ref{Hamiltonian}). If it vanishes and the DOFs are separated in
$H_{0}$, the system does not have a thermodynamic equilibrium
state. The dynamical equations are then integrable with $f$
integral invariants $J_{i}$ and the DOFs are not able to exchange
energy because of the existence of regular orbits. In such case
the closed trajectory is then only composed of nontypical
segments. Nonintegrability of the Hamiltonian is thus necessary to
achieve thermodynamic equilibrium.

As the system approaches equilibrium the segments gradually become
typical ones. This process is characterized by the entropy
production $\sigma$
\begin{equation}
\sigma \equiv \frac{dS}{dt}= k\frac{d
\ln\left(\prod_{i=1}^{f}J_{i}\right)}{dt}=-k\sum_{i=1}^{f}\frac{1}{J_{i}}\frac{\partial
H_{1}}{\partial \theta_{i}} \label{entroprod}
\end{equation}
where Eq. (\ref{Hamiltonian}) and the Hamilton equations of motion
for the action variables have been used. At equilibrium $S$
becomes $S^{(eq)}$ which does not depend on time, and the entropy
production vanishes.

Relaxation to equilibrium is described most easily through a term
$H_{1}=\sum_{i=1}^{f}(J_{i}-J_{i}^{(eq)})\theta_{i}/\tau_{r}$ in
Eq.(\ref{Hamiltonian}), with $J_{i}^{(eq)}\equiv
J_{i}(\varepsilon^{(eq)})$. This term can be thought to appear as
a consequence of collisions (or other complex interactions) that
thermalize the system. The Hamilton equation
$\dot{J}_{i}=-\partial H_{1}/\partial
\theta_{i}=-(J_{i}-J_{i}^{(eq)})/\tau_{r}$, can be integrated as
$J_{i}(t)=(J_{i0}-J_{i}^{(eq)})e^{-t/\tau_{r}}+J_{i}^{(eq)}$. From
$J_{i}= c_{i} \varepsilon_{i}^{p}$ we have that the energy
$\varepsilon_{i}$ relaxes to $\varepsilon^{(eq)}$ when $J_{i}$
relaxes to $J_{i}^{(eq)}$, and $S$ relaxes then to $S^{(eq)}$. The
entropy production for this $H_{1}$ can be calculated by means of
Eq. (\ref{entroprod}) as $\sigma =
\frac{k}{\tau_{r}}e^{-t/\tau_{r}}\sum_{i=1}^{f}(J_{i}^{(eq)}-J_{i0})/J_{i}(t)$.
This is a positive definite function which vanishes when
equilibrium is attained.

Although relaxation processes can be much more complex,
action-orbit coupling suffices to describe thermalization in the
approach to equilibrium. A relaxation process to a nonequilibrium
steady state is illustrated in section \ref{KuramotoSection}.

\section{Examples}

In this section we consider Hamiltonians which are either
integrable or made integrable through suitable approximations
based on the dynamics (e.g. the globally chaotic system). When
equilibrium quantities are mentioned it should be understood that
there is an additional mechanism in the Hamiltonian producing
relaxation to equilibrium analogous to the one described in the
previous section. Here we are only concerned with the integrable
part of the Hamiltonian, from which the relevant action-angle
variables are obtained.

\subsection{Noninteracting oscillators}

Consider a system of $N$ noninteracting harmonic oscillators
($f=N$) with Hamiltonian
$H=\sum_{i=1}^{N}\left(p_{i}^{2}/2m+m\omega_{i}^{2}q_{i}^{2}/2\right)$,
the $\varepsilon_{i}$ are related to the amplitude $A_{i}$ of each
oscillator as $\varepsilon_{i}=m\omega_{i}^{2}A_{i}^{2}/2$. The
energies $\varepsilon_{i}$ depend on the initial conditions in
which the system is prepared. The action variables are
$J_{i}=\varepsilon_{i}/\omega_{i}$ and hence $p=1$.
Eq.(\ref{superentropy}) provides the entropy as $S=k \ln (
\prod_{i=1}^{N}\varepsilon_{i}/\omega_{i})$. The equilibrium
entropy is $S^{(eq)}=k \ln (
\prod_{i=1}^{N}\varepsilon^{(eq)}/\omega_{i})$ and the total
energy is given by Eq.(\ref{energyeqi}) as $E=NkT^{(eq)}$.

\subsection{Classical ideal gas}

The Hamiltonian for a classical ideal gas of $N$ particles
($f=3N$) in a cubic box of volume $V=L^{3}$ closed by adiabatic,
rigid walls is
\begin{equation}
H=\sum_{i=1}^{3N}\left[\frac{p_{i}^{2}}{2m}+C \Theta
\left(\left|\frac{2q_{i}}{L}\right|-1\right)\right]
\end{equation}
where $C$ is an infinitely large constant and $\Theta$ is the
Heaviside step-function. The Hill's region for each DOF is just a
segment with length $L$ and the action variables are
$J_{i}=2L\sqrt{2m\varepsilon_{i}}$, so that $p=1/2$. The entropy
can be easily calculated and the equilibrium entropy is obtained
from it as $S^{(eq)}=k\ln \left [V^{N}
\left(2mE/3N\right)^{3N/2}\right]$ which coincides with the
microcanonical one, except for an undetermined constant $a$ like
in Eq. (\ref{entropy}). From Eq.(\ref{energyeqi}) the energy is
$E=3NkT^{(eq)}/2$.

We can group the $J_{i}$ appearing in the
entropy in blocks of three factors corresponding to each particle.
Permutations of the blocks cannot be distinguished and, since
there are $N!$ of these permutations the constant $a$ should be
$\propto 1/N!$. These considerations constitute the explanation of
the Gibbs' paradox. The dependence of the
equilibrium entropy on extensive variables $(N,V,E)$ match the
traditional calculations and other thermodynamic variables such as
pressure and chemical potential can be defined from this entropy
\cite{GarciaMorales06}.

\subsection{Planetary system}

Let us consider now a system of $N$ particles of mass $m$ moving
in closed orbits on the same plane subject to a gravitational
potential ($f=2N$). Neglecting the interparticle interactions, the
Hamiltonian is
\begin{equation}
H=\sum_{j}^{N}\left[\frac{1}{2m}\left(p_{r,j}^{2}+\frac{p_{\theta,j}^{2}}{r_{j}^{2}}\right)-\frac{\alpha
}{r_{j}}\right]
\end{equation}
 where subindex $p_{r,j}$, $p_{\theta,j}$ and
$r_{j}$ are the radial and angular momenta and the position of the
particle $j$, respectively. Each particle has two DOFs of
degenerate frequencies and their energies cannot be separated. As
a result, the product of actions in Eq.(\ref{superentropy}) run
now over particles instead of DOFs. The action per particle is
$J_{j}=J_{r,j}+J_{\theta,j}=\alpha
\sqrt{\frac{m}{2|\varepsilon|_{j}}}$ \cite{Landau82}. In this case
$p=-1/2$, the entropy is $S=2k\ln
\left[\prod_{j=1}^{N}\alpha\sqrt{\frac{m}{2|\varepsilon_{j}|}}\right]$
and the equilibrium entropy $S^{(eq)}=2Nk\ln
\left[\alpha\sqrt{\frac{Nm}{2|E|}}\right]$. It is to be noted that
this equilibrium is unstable. The energy is $E=-NkT^{(eq)}$ and
the heat capacity is negative, $C=-Nk$. Negative heat capacities
arise frequently in bound gravitational systems \cite{Gross}.

\subsection{Globally chaotic system in the adiabatic approximation}

Let us consider next a system with $f=2$ and
$H=\left(p_{x}^{2}+p_{y}^{2}+x^{2}y^{2}\right)/2$. It constitutes
a simple model of the classical Yang-Mills field.  The energies
$\varepsilon_{i}$ are not separated in the Hamiltonian due to the
coupling term $x^{2}y^{2}/2$. However, the Hamiltonian can be
calculated in action-angle variables by means of the adiabatic
approximation as
$H=\left(\frac{3\pi}{4\sqrt{2}}\right)^{2/3}J_{x}^{2/3}J_{y}^{2/3}$\cite{Martens89}.
The entropy is then $S=S^{eq}=k\ln
\left[\left(\frac{4\sqrt{2}}{3\pi}\right)E^{3/2}\right]$ and the
energy is $E=3kT^{(eq)}/2$. Hence, although the Hamiltonian of
this globally chaotic system is complex, the adiabatic
approximation \emph{based on the dynamics} \cite{Martens89} allows
us to write it as in Eq. (\ref{Hamiltonian}). Moreover, the
coupling $J_{x}^{2/3}J_{y}^{2/3}$ between the action variables
makes the system to be already thermalized (action variables are
not separated in the Hamiltonian) so that under the adiabatic
approximation the system is always ``at equilibrium".

\subsection{Calogero Hamiltonians with short range forces}

Finally, we consider a system of $N$ particles in one dimension
($f=N$) with nearest-neighbor interactions, periodic boundary
conditions and Hamiltonian
\begin{eqnarray}
H=&&N\lambda\sum_{i}^{N}p_{i}+\mu\sum_{i=0}^{N-1}\sqrt{p_{i}p_{i+1}}\cos\left[\nu
\left(q_{i}-q_{i+1}\right)\right] \nonumber \\
&&+\mu\sigma\sqrt{p_{i}p_{i+1}}\cos\left[\nu
\left(q_{1}-q_{N}\right)\right]
\end{eqnarray}
This Hamiltonian belongs to a class of fully integrable Calogero
Hamiltonians and can be written in action-angle variables as
$H=\sum_{i}^{N}J_{i}\omega_{i}$ where the $\omega_{i}$'s are the
eigenvalues of the matrix \cite{Karimipour97}
\begin{equation}
C=\left(%
\begin{array}{ccccc}
  0 & 1 & . & . & \sigma \\
  1 & 0 & 1 & . & . \\
  . & 1 & 0 & 1 & . \\
  . & . & 1 & 0 & 1 \\
  \sigma & . & . & 1 & 0 \\
\end{array}%
\right)
\end{equation}
where a dot stands for $0$. If $C$ is any hermitian matrix, one
can construct a wide class of interesting Hamiltonians that can be
used to model many-particle interacting systems with short- and
long-range forces \cite{Karimipour97}. The entropy, equilibrium
entropy and energy, $S=k \ln (
\prod_{i}^{N}\varepsilon_{i=1}/\omega_{i})$, $S^{(eq)}=k \ln
\left[\left(E/N\right)^{N}\prod_{i=1}^{N}1/\omega_{i}\right]$ and
$E=NkT^{(eq)}$, coincide with that of $N$ \emph{noninteracting}
harmonic oscillators with frequencies $\omega_{i}$.

\section{Nonequilibrium thermodynamics: Entropy production in a network of coupled oscillators}
\label{KuramotoSection}

The Kuramoto model \cite{Kuramoto} (see also \cite{Acebron} and
\cite{Strogatz} for excellent reviews) has received great recent
interest because it provides a simple model for synchronization.
This model considers a network of oscillators which are coupled
harmonically in phase. After a transient, and under certain
conditions as explained below, a fraction of the oscillators in
the network become synchronized. This means that both the
effective frequency of the oscillators and the phase becomes
equal, in spite of the fact that they have different natural
frequencies and they start with very different phases
\cite{Acebron}.

The Kuramoto model is a particular case of the so-called phase
models \cite{Hoppensteadt,Hudson} in which angles are coupled
through a harmonic function. Every dynamics yielding oscillations
to a limit cycle can be transformed in a phase model by means, for
example, of the Hilbert transform \cite{Hudson}. Phase models are
naturally linked to action-angle descriptions: the phase of each
oscillator runs periodically on the limit cycle confining the
trajectory, and in completing each cycle a mechanical action
(which is conjugate to the phase) is carried out by the system. In
our view, the action variables control the thermodynamic
properties of the system through Eq. (\ref{superentropy}) and this
opens the possibility of studying these properties using our
formalism.

\begin{figure}
\begin{center}
\includegraphics[angle=0, width=0.4\textwidth]{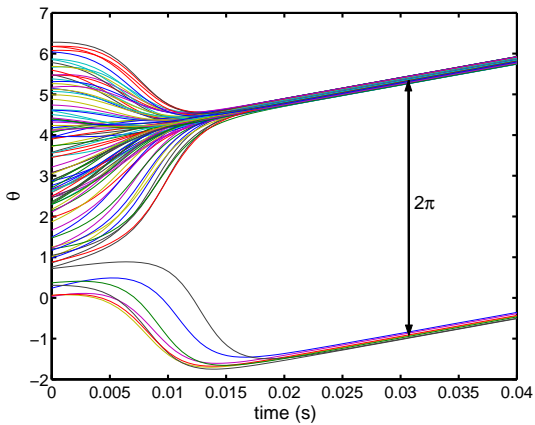}
\caption{Time evolution of the phases of 100 oscillators for
strong coupling K=5 obtained by numerical integration of Eqs.
(\ref{Kuramoto}). The system starts with a random distribution of
phases and after a short transient all oscillators become
synchronized with the emergence of two branches with the same
effective frequency.} \label{Fig2}
\end{center}
\end{figure}

The action-angle Hamiltonian yielding the Kuramoto dynamics is
\begin{equation}
H=\sum_{i}^{N}J_{i}\omega_{i}^{\circ}+K\sum_{j \ne
i}^{N}\sum_{i}^{N}J_{i}\sin(\theta_{j}-\theta_{i})
\end{equation}
The first term on the r.h.s. describes the uncoupled system of
oscillators with natural frequencies $\omega_{i}^{\circ}$ while
the second depends on a global coupling parameter $K$ that can be
arbitrarily large. The system is nonintegrable because of the
explicit appearance of the angle variables in the second term.
This makes every action variable $J_{i}$ to evolve in time. The
Hamilton equations of motion are
\begin{eqnarray}
\omega_{i}&=&\frac{\partial H}{\partial
J_{i}}=\omega_{i}^{\circ}+K\sum_{j \ne i}^{N}\sin(\theta_{j}-\theta_{i}) \label{Kuramoto}\\
\dot{J}_{i}&=&-\frac{\partial H}{\partial
\theta_{i}}=KJ_{i}\sum_{j \ne i}^{N}\cos(\theta_{j}-\theta_{i})
\label{Ji}
\end{eqnarray}
Eqs. (\ref{Kuramoto}) constitutes the Kuramoto dynamics. It
establishes the effective frequency of the $i$-th oscillator,
which is driven in phase through the all-to-all coupling. In the
limit of vanishing coupling $K$ each DOF oscillates with its
natural frequency and the oscillators are not synchronized. For
$K$ higher than a critical value $K_{c}$, however, some
oscillators become synchronized after a transient. In the limit of
strong coupling all oscillators are synchronized. This situation
is depicted in Fig.\ref{Fig2}, which has been obtained by
numerical integration of Eqs. (\ref{Kuramoto}) by employing a
fourth order Runge-Kutta algorithm for $N=100$ oscillators
starting at random initial phases and with a uniform distribution
of natural frequencies in the range (0, N] rad/s. The phases of
the oscillators converge into two branches with the same effective
frequency and phase. The suitability of the action-angle variables
for the parameters considered was confirmed in the simulations by
checking Eq. (\ref{crit2}) at each time.

\begin{figure}
\begin{center}
\includegraphics[angle=0, width=0.4\textwidth]{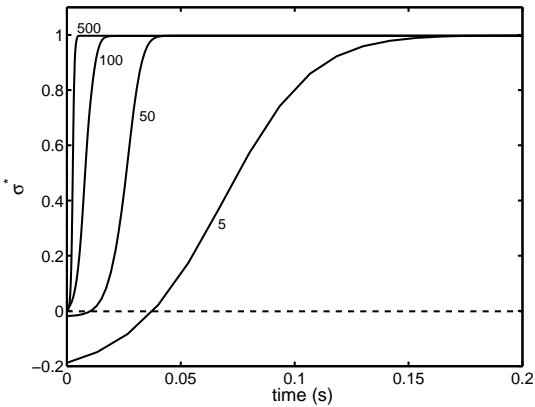}
\caption{Time evolution of the normalized entropy production
$\sigma^{*}=\sigma/[kKN(N-1)]$ for different networks of
oscillators with the values of $N$ indicated in the figure and
coupling parameter $K=5$. The system starts with a random
distribution of phases and after a short transient all oscillators
become synchronized.} \label{Fig3}
\end{center}
\end{figure}

Eqs. (\ref{Ji}) represent the dynamics of the conjugate action
variables and the entropy production can be calculated from
Eq.(\ref{entroprod}) as
\begin{equation}
\frac{\sigma}{k}= K\sum_{j \ne
i}^{N}\sum_{i}^{N}\cos(\theta_{j}-\theta_{i}) \label{sigma}
\end{equation}
Remarkably, this equation does not depend on the $J_{i}$ and
$\sigma$ can be evaluated after numerical integration of
Eqs.(\ref{Kuramoto}). If we consider a high number of oscillators,
the r.h.s of Eq.(\ref{sigma}) vanishes when these are not
synchronized, since the phases are distributed uniformly. In the
limit of high coupling all oscillators are synchronized, the
phases of some oscillators become equal and the r.h.s of
Eq.(\ref{sigma}) becomes approximately $\sim K N^{2}$ attaining a
finite and positive value. This positive sign of the entropy
production for networks with many oscillators and long times is
perfectly consistent with the Second Law of Thermodynamics. In
Fig. \ref{Fig3} we show the value of the normalized entropy
production $\sigma^{*} \equiv \sigma/[kKN(N-1)]$ for different
network sizes and in the strong coupling limit ($K=5$). The
initial phases are chosen at random. It is observed that in the
nonequilibrium steady state, when the oscillators become
synchronized, this entropy production is positive and finite in
all cases. For networks with high numbers of oscillators, the
entropy production is positive at all times in all trajectories
starting at random phases. For low number of oscillators and small
time scales (far from the thermodynamic limit), negative entropy
production can be observed.

\begin{figure}
\begin{center}
\includegraphics[angle=0, width=0.4\textwidth]{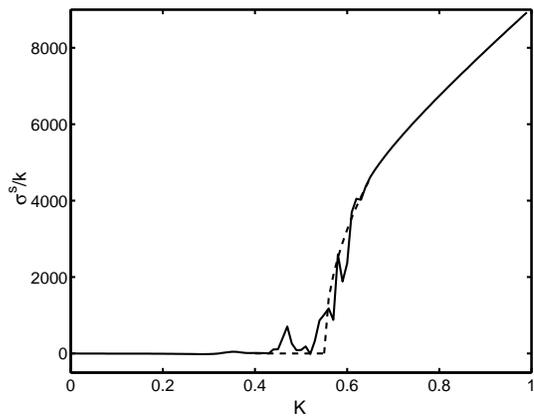}
\caption{Entropy production in the long time limit as a function
of the coupling parameter for a network of $N=100$ oscillators.
The entropy production vanishes for coupling parameter lower than
a critical value around $K=0.55$. At the onset of synchronization,
the entropy production becomes positive and finite. In the limit
of strong coupling, all oscillators are synchronized and the
entropy production approaches a line with slope $\sim N^{2}$.}
\label{Fig4}
\end{center}
\end{figure}

Another interesting issue concerns the characterization of
disorder. The nonequilibrium entropy calculated above is directly
related to the activity of the network of oscillators and acts
itself as an order parameter of the system. In Fig. (\ref{Fig4})
we show the value of the entropy production at the steady state
$\sigma^{s} \equiv \sigma(t \to \infty)$ as a function of the
coupling parameter for a network of $N=100$ oscillators. Quite
interestingly, the entropy production is zero for low values of
the coupling parameter in which the oscillators are not
synchronized. Then, at a critical value $K_{c}$ around 0.55, the
entropy production becomes increasingly positive with increasing
coupling and tends asymptotically to a line with slope $\sim
N^{2}$ at strong coupling. This finite, positive entropy
production arises exactly when a fraction of oscillators in the
network gets synchronized through the coupling. It represents the
price that we have to pay for getting this ordered phase
stabilized in time. For coupling parameter below the critical
value, the network is disordered and there is no dissipation (i.e.
the entropy production vanishes).

When all natural frequencies of the oscillators are equal we can
prove analytically that the entropy production increases in
driving the system to the stationary state. In this case we have
$\omega_{i}^{\circ}=\omega^{\circ}$ for all $i$, and we can change
the angle variables to $\phi_{i}=\theta_{i}-\omega^{\circ}t$. Eqs.
(\ref{Kuramoto}) read in this case as
\begin{equation}
\omega_{i}=\frac{\partial H}{\partial
J_{i}}=K\sum_{j \ne i}^{N}\sin(\phi_{j}-\phi_{i}) \label{Kuramoto2}\\
\end{equation}
and the entropy production, Eq. (\ref{sigma}), is
\begin{equation}
\frac{\sigma}{k}= K\sum_{j \ne
i}^{N}\sum_{i}^{N}\cos(\phi_{j}-\phi_{i}) \label{sigma2}
\end{equation}
By comparing Eqs.(\ref{Kuramoto2}) and (\ref{sigma2}) we observe
that
\begin{equation}
\omega_{i}=\frac{1}{k}\left(\frac{\partial \sigma}{\partial
\phi_{i}}\right)
\end{equation}
and therefore
\begin{equation}
\frac{d\sigma}{dt}=
\sum_{i}^{N}\left(\frac{\partial\sigma}{\partial
\phi_{i}}\right)\omega_{i}=\frac{1}{k}\sum_{i}^{N}\left(\frac{\partial\sigma}{\partial
\phi_{i}}\right)^{2} \ge 0 \label{sigmarat1}
\end{equation}
which is the result that we wanted to prove.

\section{Conclusions and perspectives}

We have shown that thermodynamics can be grounded on classical
mechanics with the help of the Maupertuis principle. In our
approach, thermal states are fully characterized by appropriate
action variables. The nonintegrable part of the Hamiltonian
dictates relaxation to equilibrium, an evolution driven to an
steady state or a departure from equilibrium. This approach is
consistent with the Second Law of Thermodynamics for macroscopic
systems. The results obtained are, therefore, quite general and
can be applied to systems with arbitrary size both in and out of
equilibrium. Although we do not make explicit use of probability,
it is also shown that the latter can be introduced from classical
mechanics yielding an Einstein relationship at the thermodynamic
limit. The main difficulty in our approach comes from finding the
appropriate set of action-angle variables from a given
Hamiltonian. We have shown, however, that in many relevant cases
this is possible provided that a criterion on the suitability of
these variables (coming from demanding a necessary consistency for
the definition of the nonequilibrium entropy) is satisfied. Our
approach has then been illustrated by applying it to systems with
very different physical properties (ideal and globally chaotic
systems and many-body Hamiltonians with short and long-range
interactions). The dynamical definitions of thermodynamic
quantities reproduce previous results derived from statistical
mechanics. Finally, the relevance and applicability of our ideas
to nonequilibrium situations has been illustrated by evaluating
the (dynamically-defined) entropy production of a network of
coupled oscillators (Kuramoto model). This quantity proves to be
physically meaningful giving a measure of dynamical order. We find
that its positive value at the steady state is directly related to
the emergence of an ordered, synchronized phase from the
collective dynamics of the coupled oscillators. This opens the
possibility of studying the thermodynamics of pattern recognition
\cite{Hoppensteadt,Hoppensteadt2}.

We acknowledge fruitful conversations with Profs. Katharina
Krischer, Miguel A. Sanchis-Lozano and Alejandro Casanovas. This
work was supported by the MEC (Ministry of Education and Science
of Spain) and FEDER under Project No. MAT2005-01441.

\section*{APPENDIX A: ERGODICITY}

The thermodynamic formalism presented in this paper does not require ergodicity. However, it
 embodies ergodicity when Birkhoff's theorem holds.
From this theorem (see \cite{Arnold78} p. 286), time averages
$<...>_{\tau_{c}}$ coincide with phase space averages
$<...>_{\Gamma}$ when all frequencies $\omega_{i}=\partial H
/\partial J_{i}$ are independent. In this case, Eq.
(\ref{energyeqi}) becomes the statement of the equipartition
principle and the system is ergodic. We then have
$fkT^{(eq)}=<2K>_{\tau_{c}}=<2K>_{\Gamma}=f\Phi(E)/\Omega(E)$
\cite{Khinchin} where $\Phi=\int d \Gamma \Theta
\left[E-H(\textbf{q},\textbf{p})\right]$ and $\Omega= \int d
\Gamma \delta \left[E-H(\textbf{q},\textbf{p})\right]$. $\Theta$
and $\delta$ are the Heaviside and the Dirac delta functions,
respectively, and $d\Gamma$ is the phase space volume element.
Since $\Omega= \partial \Phi/\partial E$, we find that
$fkT^{(eq)}=f\left(\partial \ln \Phi/\partial E\right)^{-1}$.
Taking this expression to Eq. (\ref{entropydif}) we obtain
$(J_{c})^{f}=\Phi$ and $S^{(eq)}=k\ln \Phi$. This is the so-called
(Hertz) volume entropy which is an adiabatic invariant
\cite{Adib04, Berdichevsky}. Birkhoff theorem does not hold,
however, when there exists any integral relation between the
frequencies $\omega_{i}$ \cite{Arnold78}. In these cases, Eq.
(\ref{energyeqi}) can depart from ergodicity and equipartition.

\section*{APPENDIX B: THE MIXING OF TWO IDEAL FLUIDS}

\begin{figure}
\begin{center}
\includegraphics[angle=0, width=0.4\textwidth]{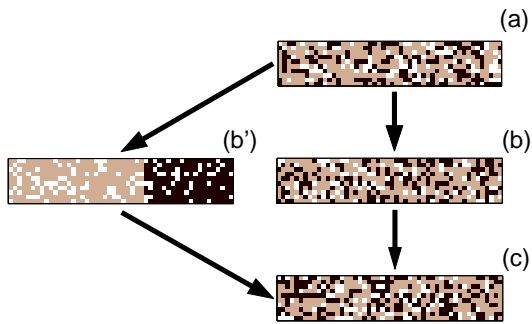}
\caption{Two possible trajectories joining the disordered
configurations (a) and (c) of two mixed fluids. The trajectory
going through (b') is forbidden by the Maupertuis principle since
it is possible to find disordered configurations (b) between (a)
and (c) so that the abbreviated action ($\equiv$ the length of the
trajectory between (a) and (c) in the 3N-dimensional configuration
space) is a minimum.} \label{Fig5}
\end{center}
\end{figure}

A popular topic in equilibrium thermodynamics is the mixing of two
ideal fluids (see \cite{Lebowitzb} for a beautiful explanation of
Boltzmann ideas). Let us consider two ideal fluids at equilibrium
mixed in a container closed by adiabatic rigid walls. The
particles interact only through elastic collisions. This isolated
system is depicted schematically in Fig. \ref{Fig5} (a) and is
composed of pink and black particles. We ask now why we do not
observe an spontaneous departure from this homogeneous mixture
leading to a separation of both fluids as represented in Fig.
\ref{Fig5} (b').

The Maupertuis principle requires that
\begin{equation}
\delta J_{E}=\int_{\textbf{q}_{0}}^{\textbf{q}_{1}}
\textbf{p}d\textbf{q} =
\sqrt{2mE}\delta\int_{\textbf{q}_{0}}^{\textbf{q}_{1}}ds=0
\end{equation}
where $ds$ is the differential of arc length on the trajectory.
The last equality states that the trajectory of the motion between
two points in the configuration space (say (a) and (c) in Fig.
\ref{Fig5}) is the one with \emph{minimal length}. We now can
immediately realize why we do not observe (b') starting from (a).
\emph{The number of points in the trajectory in which the system
looks macroscopically disordered as (a), (b) and (c) is large and
points (a) and (c) are connected through points as (b) and not as
(b'), because going through intermediate states of disorder as (b)
extremizes the abbreviated action}. It can be intuitively seen
with the help of Fig. \ref{Fig5} that, from a mixed state (a),
separating pink particles from black particles (b') and then
mixing them again (c) leads to a longer trajectory than passing
through intermediate mixed states such as (b). Attaining (b') from
(a) requires each particle to travel a certain distance so that it
is then located in the appropriate macroscopic subvolume. This
requirement is not necessary if the trajectory passes through
intermediate disordered states such as (b). Each individual
particle travels now a much shorter distance and so is the total
length of the trajectory in the 3N-space. By Maupertuis principle,
the shortest path is the one chosen by the motion (in the typical
time interval of a measurement) so that passing through (b') is
forbidden.

Although points like (b') lie far apart of the trajectory of the
motion, the above considerations do not rule out totally the
possibility of attaining them. Some of these points can belong to
the whole closed trajectory \emph{but lie in segments which have a
total duration which is very short compared to the whole closed
trajectory}. These segments \emph{are not} the typical ones of the
trajectory. The typical segments, whose properties contribute most
significantly to the time average on the closed trajectory join
points as (a)-(b)-(c).

All above considerations can be easily extended to any Hamiltonian
system at equilibrium but then the quantity equivalent to the
minimum of the abbreviated action is not the minimal geometric
length of the trajectory but the extremum of the quantity
\begin{equation}
\int_{\textbf{q}_{0}}^{\textbf{q}_{1}}\sqrt{2m(E-U)}ds
\end{equation}
The integrand $\sqrt{2m(E-U)}$ plays here a similar role to the
index of refraction in Fermat's principle of the minimum optical path.

\end{document}